\documentclass[letterpaper, 10 pt, conference]{ieeeconf}

\IEEEoverridecommandlockouts                
\usepackage{mathtools,amssymb,commath}
\usepackage{amsmath}
\usepackage{graphicx,import}
\usepackage{verbatim}
\usepackage{color}
\usepackage{hyperref}
\hypersetup{colorlinks=true}  
\usepackage{subcaption}

\DeclareMathOperator*{\st}{\mathbf{s.t.}\,}					

\usepackage{mathtools,amsthm}

\title{\LARGE \bf Stochastic and Robust MPC for Bipedal Locomotion: A Comparative Study on Robustness and Performance}

\author{Ahmad Gazar$^{1}$, Majid Khadiv$^{1}$, Andrea Del Prete$^{2}$, Ludovic Righetti$^{1,3}$ 
\thanks{
This work was partially supported by the European Unions Horizon 2020 research and innovation program under Grant Agreement 780684, the European Research Councils under Grant 637935 and the National Science Foundation under Grant CMMI-1825993.
The authors thank the International Max Planck Research School for Intelligent Systems (IMPRS-IS)
for the non-financial support of Ahmad Gazar. }
\thanks{$^{1}$ Max Planck Institute for Intelligent Systems, Tuebingen, Germany. {\tt\small firstname.lastname@tuebingen.mpg.de}}%
\thanks{$^{2}$ Industrial Engineering Department, University of Trento, Italy. {\tt\small andrea.delprete@unitn.it}}%
\thanks{$^{3}$ Tandon School of Engineering, New York University, New York, USA. {\tt\small ludovic.righetti@nyu.edu}}%
}

\begin{document}

\maketitle
\thispagestyle{empty}
\pagestyle{empty}

\begin{abstract}

Linear Model Predictive Control (MPC) has been successfully used for generating feasible walking motions for humanoid robots. However, the effect of uncertainties on constraints satisfaction has only been studied using Robust MPC (RMPC) approaches, which account for the worst-case realization of bounded disturbances at each time instant. In this letter, we propose for the first time to use linear stochastic MPC (SMPC) to account for uncertainties in bipedal walking. 
We show that SMPC offers more flexibility to the user (or a high level decision maker) by tolerating small (user-defined) probabilities of constraint violation. Therefore, SMPC can be tuned to achieve a constraint satisfaction probability that is arbitrarily close to 100\%, but without sacrificing performance as much as tube-based RMPC. 
We compare SMPC against RMPC in terms of robustness (constraint satisfaction) and performance (optimality). 
Our results highlight the benefits of SMPC and its interest for the robotics community as a powerful mathematical tool for dealing with uncertainties. 

\end{abstract}

\section{Introduction}
\label{sec:intro}


Control of humanoid robots is challenging due to limiting constraints on contact forces, and nonlinear switching dynamics. Furthermore, guaranteeing safety for humanoids is critical, as collision with the environment or falling down can cause severe damage to the robot and its surroundings. 
Linear MPC \cite{mayne2007}\cite{rawlings} is a powerful tool for designing real-time feedback controllers subject to state and input constraints, which makes it a prime candidate for generating a wide range of feasible reference walking motions for humanoid robots  \cite{wieber2016,herdt2010,sherikov2014whole}. However, the theoretical guarantees 
associated with MPC (e.g., constraint satisfaction guarantees) can easily be lost due to external disturbances or the discrepancy between the nonlinear dynamics of the robot and the linearized model used in control.

Recently, \cite{brasseur2015robust,dai2016planning} studied how to account for the bounded error in constraint satisfaction due to the approximation of the nonlinear center of mass (CoM) dynamics, and \cite{bohorquez2017adaptive} investigated nonlinear constraints due to step timing adaptation. However,  $1)$ they do not account for the closed-loop tracking errors due to disturbances,
$2)$ there are no robustness guarantees of constraints satisfaction in the presence of different disturbances, which is critical for generating safe walking motions.

Linear Robust MPC (RMPC) schemes have been extensively studied in the control literature \cite{mayne2001,chisci2001,mayne2005}.
Recently, \cite{villa2017} used the well-known \textit{tube-based RMPC}  approach originally developed in \cite{mayne2005} for generating robust walking motions for humanoid robots, taking into account the effects of additive compact polytopic uncertainties on the dynamics. Using a state feedback control policy and a pre-stabilizing choice of static dead-beat gains, they showed that constraints are guaranteed to be satisfied for all disturbance realizations inside the disturbance set. 
A drawback of RMPC is that the constraints are designed to accommodate for the worst-case disturbance, which is quite conservative and sacrifices performance (optimality) to guarantee hard constraints satisfaction.   

In order to relax the conservativeness of RMPC, SMPC \cite{cannon2016, mesbah2017, scattolini2016, Lorenzen2017} exploits the underlying probability distribution of the disturbance realizations. Furthermore, SMPC offers a flexible framework by accounting for \textit{chance constraints}, where constraints are expected to be satisfied within a desired probability level. Depending on how critical the task is, the user can tune the desired probability level between the two extremes of \textit{almost} hard constraint satisfaction (as in RMPC) and complete negligence of disturbances (as in nominal MPC). This flexibility becomes very practical, since a humanoid robot needs to move in dynamic environments where some of the constraints can be more critical than others. For example, moving through a narrow doorway or walking in a crowd \cite{ciocca2019effect}, the robot needs to reduce the sway motion of its CoM to reduce the probability of collision. However, for walking on challenging terrains with partial footholds \cite{wiedebach2016walking}, the robot has to bring the foot center of pressure (CoP) as close as possible to the center of the contact area. Many other tasks can be considered somewhere between those situations. To this end, SMPC can be a powerful and systematic tool for dealing with constraint satisfaction in different environments and tasks. Moreover, small errors are typically more likely to occur in practice. It might therefore be more appropriate to explicitly consider the distribution of disturbances instead of treating all of them equally as in RMPC, which often lead to a conservative behavior. 

In this letter, we revisit the problem of generating reference walking motions for humanoid robots using a linear inverted pendulum model (LIPM) subject to additive uncertainties on the model. Our contributions are the following:
\begin{itemize}
\item We introduce linear SMPC to generate stable walking, taking into account stochastic model uncertainty subject to individual chance constraints. 
\item We analyze the robustness of SMPC to worst-case disturbances, drawing an interesting connection between robust and stochastic MPC, and highlighting their fundamental difference.
\item We compare SMPC, RMPC, and nominal MPC in terms of robustness (constraints satisfaction) and performance. Our tests focus on stochastic bounded disturbances (generated with a truncated Gaussian distribution), which are a good approximation of real disturbances, such as joint torque tracking errors~\cite{delprete2016}. We empirically show that SMPC can achieve hard constraint satisfaction, while being significantly less conservative than RMPC.
\end{itemize}

\section{Background}
\label{sec:background}

\subsection{Notation}
\label{subsec:notation}

\begin{itemize}
  \item $x_t$ represents the value of $x$ at time $t$, while $x_{t+i|t}$ denotes the value of $x$ at the future time $t+i$ predicted at time $t$  
  
  \item $\mathcal{A} \oplus \mathcal{B}=\{a+b \,|\, a\in \mathcal{A},\, b\in \mathcal{B}\}$ refers to the Minkowski set sum 
  
  \item $\mathcal{A} \ominus \mathcal{B}=\{a \in \mathcal{A} \,|\, a+b \in \mathcal{A}, \, \forall b \in \mathcal{B}\}$ refers to the Pontryagin set difference 
  
  \item a random variable $x$ following a distribution $\mathcal{Q}$ is denoted as $x \sim \mathcal{Q}$, and $\mathbb{E}[x]$ is the expected value of $x$  
\end{itemize}

\subsection{Linear model of walking robots}
The dynamics of the CoM of a walking robot, under the assumption of rigid contacts with a flat ground, can be modelled as follows~\cite{tedrake2016}:
\begin{IEEEeqnarray}{LLL}
\label{eq:centroidal dynamics}
p^{x,y} = c^{x,y}-\frac{m_{tot} \, c^z \,\ddot{c}^{x,y}+S\dot{L}^{x,y}}{m_{tot} (\ddot{c}^z + g^z)}, \IEEEyesnumber 
\end{IEEEeqnarray}
where $c \in \mathbb{R}$ denotes the CoM position in the lateral directions of motion $^{x,y}$. The total mass of the robot is denoted by $m_{tot}$, the matrix $S = \begin{bsmallmatrix}0 & -1  \\1&0\end{bsmallmatrix}$ is a rotation matrix, with the center of pressure (CoP) $p \in \mathbb{R}$ being constrained inside the convex hull of the contact points $\mathcal{U}$
\begin{IEEEeqnarray}{LLL}
p^{x,y} \in \mathcal{U}. \IEEEyesnumber
\end{IEEEeqnarray}
Under the assumption of constant CoM height $c^z$ and constant angular momentum $L$, the dynamics~\eqref{eq:centroidal dynamics} can be simplified to the well-known Linear Inverted Pendulum Model (LIPM), resulting in the following  linear relationship between the CoM and the CoP
\begin{IEEEeqnarray}{LLL} 
\label{eq:LIPM}
\ddot{c}^{x,y} = \omega_n^2(c^{x,y}- p^{x,y}),
\end{IEEEeqnarray}
where $\omega_n = \sqrt{\frac{g^z}{c^z}}$ represents the system's natural frequency, and $g^z$ being the norm of the gravity vector along $z$. From now on, we will drop $^{x,y}$ superscripts for convenience. 
\subsection{Nominal linear MPC for bipedal locomotion}
\label{sub-section:deterministic MPC for LIPM}
Consider the discrete-LTI dynamics (\ref{eq:LIPM}) subject to state and control constraints: 
\begin{IEEEeqnarray}{LLL} 
\IEEEyesnumber
x_{t+i+1} = Ax_{t+i} + Bu_{t+i},  \IEEEyessubnumber \label{eq:nominal dynamics}\\
x_{t+i+1} \in \mathcal{X}, \qquad 
u_{t+i} \in \mathcal{U}, \IEEEyessubnumber
\end{IEEEeqnarray}
where the state $x = \begin{bmatrix}c & \dot{c}\end{bmatrix}^\top \in \mathbb{R}^n$, with $n=2$, and the control input $u = p \in \mathbb{R}^m$, with $m=1$. $\mathcal{X}$ represents the set of linear kinematic constraints of the robot, like self collision, maximum stride length, etc. MPC deals with solving the following optimal control problem (OCP) at every sampling time $t$:
\begin{IEEEeqnarray}{LCL}
\IEEEyesnumber
\label{nominal mpc}  
\min_{\mathbf{u}}\quad J_N(x_t,\mathbf{u})  
\IEEEyessubnumber
\label{eq:determinstic cost function}\\
\st \nonumber \\
\quad\quad x_{t+i+1|t} =  Ax_{t+i|t} + Bu_{t+i|t},  \IEEEyessubnumber\\
\quad\quad x_{t+i+1|t} \in \mathcal{X},  \IEEEyessubnumber\\
\quad\quad u_{t+i|t} \in \mathcal{U},  \IEEEyessubnumber\\
\quad\quad x_{t|t} = x_t ,  \IEEEyessubnumber \\
\quad\quad i = 0,...,N-1.\IEEEyessubnumber  
\end{IEEEeqnarray}
$\mathbf{u} = \{u_{t|t}, u_{t+1|t}, ..., u_{t+N-1|t}\}$ denotes the control sequence along the prediction horizon $N$ and $\mathbf{u}^\ast(x_t)$ is the minimizer of (\ref{nominal mpc}) given the current initial condition $x_t$. The above MPC scheme applies only the first control action $u^\ast_{t|t}(x_t)$ of the optimal open-loop control sequence. We avoided using terminal constraints (e.g capturability \cite{koolen2012capturability}) in our comparison, since to the best of our knowledge there is no systematic way for handling terminal constraints in SMPC as in nominal MPC and RMPC. One of the options for generating viable reference walking trajectories using the above MPC scheme without terminal constraints is to minimize one of the CoM derivatives, adding it to the cost function $J_N$ \cite{wieber2016}\cite{herdt2010}\cite{tedrake2016}. With a sufficiently long $N$ 
a valid choice of the cost function in (\ref{eq:determinstic cost function}) can be  
\begin{IEEEeqnarray}{LLL}
\label{eq:cost_function}
J_N(x_t,\mathbf{u}) & = \sum^{N-1}_{i=0} \alpha(\dot{c}^d_t-\dot{c}_{t+i|t})^2  + \beta(c^d_t-c_{t+i|t})^2  \\ \nonumber &  + \gamma(p^d_t-p_{t+i|t})^2.
\end{IEEEeqnarray}
 $c^d_{t}$, and $\dot{c}^d_{t}$ represent desired walking direction and velocity of the robot respectively. $p^d_{t}$ denotes the desired CoP tracking position, which is usually chosen to be at the center of $\mathcal{U}$ for robustness. $\alpha, \beta$ and $\gamma$ are user-defined weights.

\newtheorem{property}{Property}
\newtheorem{assumption}{Assumption}
\newtheorem{remark}{Remark}

\section{Tube-based Robust MPC (RMPC)}
\label{sec:RMPC}
Two Tube-based linear RMPC versions were first introduced in \cite{mayne2001} and \cite{chisci2001}. We follow the approach of \cite{mayne2001} as it has been more commonly used in the control community, and recently in \cite{villa2017} for bipedal locomotion. Note however that our qualitative results and comparison with SMPC would still hold for \cite{chisci2001}.

\subsection{Robust OCP formulation and control objective}
\label{subsec:robust problem}
Consider the following discrete-LTI prediction model subject to additive stochastic disturbance $w_t$:
\begin{IEEEeqnarray}{LLL}
\IEEEyesnumber
\label{eq:robust problem setting}
x_{t+i+1|t} = Ax_{t+i|t} + Bu_{t+i|t} + w_{t+i}, \IEEEyessubnumber \label{eq:robust disturbed dynamics}\\
x_{t+i+1|t} \in \mathcal{X}, \IEEEyessubnumber 
\label{eq:robust state constraints}\\
u_{t+i|t} \in \mathcal{U}. \IEEEyessubnumber 
\label{eq:robust control constraints}
\end{IEEEeqnarray}
\begin{assumption}{(Bounded disturbance)}
$w_{t+i} \in \mathcal{W}$ for $i = 0,1,2, ...$ is a disturbance realization, with $\mathcal{W}$ denoting a polytopic compact (closed and bounded) disturbance set containing the origin in its interior.
\end{assumption}
Consider the nominal state $s_t$ evolving as
\begin{IEEEeqnarray}{LLL}
\label{eq:nominal state dynamics}
s_{t+i+1|t} = As_{t+i|t} + Bv_{t+i|t},
\end{IEEEeqnarray} 
under the control action $v_{t+i|t}$. 
The main control objective of Tube-based RMPC is to bound the evolution of the closed-loop state error $e_t = x_t - s_t$
using an auxiliary state feedback control law
\begin{IEEEeqnarray}{LLL}
\label{eq:robust control law}
\IEEEyesnumber
u_{t+i|t} = v_{t+i|t}(x_t) + K(x_{t+i|t} - s_{t+i|t}),
\end{IEEEeqnarray}
where $K \in \mathbb{R}^{m \times n}$ is a fixed pre-stabilizing feedback gain for (\ref{eq:robust disturbed dynamics}), and $v_{t+i|t}(s_t)$ is the decision variable of the MPC program. 
By subtracting (\ref{eq:nominal state dynamics}) from (\ref{eq:robust disturbed dynamics}), and applying the control law in (\ref{eq:robust control law}), the error dynamics is
 \begin{IEEEeqnarray}{LLL}
    \label{eq:robust error dynamics}
    \IEEEyesnumber
     e_{t+i+1} = A_K e_{t+i} + w_{t+i},
\end{IEEEeqnarray}
with $A_K \overset{\Delta}{=}  A+BK$ being Schur (eigenvalues inside unit circle). The propagation of the closed-loop error dynamics (\ref{eq:robust error dynamics}) converges to the bounded set
 \begin{IEEEeqnarray}{LLL}
    \label{eq:mRPI}
    \IEEEyesnumber
     \Omega = \bigoplus^{\infty}_{t=0} A_K^t \mathcal{W}.
\end{IEEEeqnarray}
Hence the limit set of all disturbed state trajectories $x_t$ lie within a neighborhood of the nominal trajectory $s_t$ known as a \textit{tube of trajectories}.
It is clear that if $\mathcal{W} = \{0\} \rightarrow \Omega= \{0\}$, and the tube of trajectories collapses to a single trajectory, which is the solution of (\ref{eq:nominal state dynamics}). In set theory, $\Omega$ is called the \textit{minimal Robust Positive Invariant} (mRPI) set, or \textit{Infinite Reachable Set}. 
We recall some standard properties of disturbance invariant sets that will be used to design tightened constraint sets in the next subsection. 
\begin{property}
\label{property: positive invariance}
Positive Invariance \\
A set $\mathcal{Z}$  
is said to be a \textit{robust positively invariant} (RPI) set \cite{blanchini2008} for the system (\ref{eq:robust disturbed dynamics}) iff 
\begin{IEEEeqnarray}{LLL}
    \label{eq:RPI}
    \IEEEyesnumber
    A_K \mathcal{Z} \oplus \mathcal{W} \subseteq \mathcal{Z},
\end{IEEEeqnarray}
\end{property}  
i.e. if $e_0 \in \mathcal{Z} \Rightarrow e_t \in \mathcal{Z} \,\,\,  \forall t \geq 0$. In simple words, once the error is driven to $\mathcal{Z}$ it will remain inside $\mathcal{Z}$ for all future time steps if subject to the bounded disturbance $w_{t+i} \in \mathcal{W}$. 
\begin{property}
Minimal Robust Positive Invariance (mRPI) \\
The mRPI set $\Omega$ (\ref{eq:mRPI})  of   (\ref{eq:robust disturbed dynamics}) is the RPI set in $\mathbb{R}^n$ that is contained in every closed RPI set of (\ref{eq:robust disturbed dynamics}).
\end{property} 
An outer-approximation of the mRPI set $\Omega$ can be computed using the approach of \cite{rakovic2005}. The size of $\Omega$ depends on the system's eigenvalues, the choice of $K$, and $\mathcal{W}$.
\subsection{State and control back-off design}
\label{subsec:robust back-offs}
Using the mRPI set $\Omega$, and the stabilizing feedback gains $K$, the state and control constraint sets are tightened as 
\begin{IEEEeqnarray}{LLL}
\IEEEyesnumber
s_{t+i+1|t} \in \mathcal{X} \ominus
\Omega, \IEEEyessubnumber
\label{eq:state back-off}\\
v_{t+i|t} \in \mathcal{U} \ominus K\Omega. \IEEEyessubnumber
\label{eq:control back-off}
\end{IEEEeqnarray}
The new tightened state and control constraint sets are often called \textit{backed-off constraints}. Satisfying the backed-off constraints (\ref{eq:state back-off})-(\ref{eq:control back-off}) using the control law (\ref{eq:robust control law}), ensures the satisfaction of \eqref{eq:robust state constraints}-\eqref{eq:robust control constraints}.
\begin{remark}
\label{remark:choice of feedback gains}
Following the choice of dead-beat pre-stabilizing feedback gains $K$ proposed in~\cite{villa2017}, we get $K\Omega =  K\mathcal{W}$, which allows us to compute $K\Omega$ exactly (whereas usually this needs to be approximated using numerical techniques). The dead-beat gains are also a practical choice, since they lead to the smallest control back-off $K\Omega$~\cite{villa2017}.    
\end{remark}{}
\subsection{Tube-based RMPC algorithm}
\label{subsec:RMPC algorithm}
The tube-based RMPC scheme solves the OCP in (\ref{eq:robust problem setting}) by splitting it into two layers;
\begin{enumerate}
    \item MPC layer: computes feasible feedfoward reference control actions $\mathbf{v}^\ast(s_t)$ every MPC sampling time $t$ subject to the backed-off state and control constraints as follows
\begin{IEEEeqnarray}{LCL}
\label{eq:RMPC}  \IEEEyesnumber
\min_{\mathbf{v}} \quad  J_N(s_t,\mathbf{v}) = (\ref{eq:cost_function}) \IEEEyessubnumber\\
\st \nonumber \\ \quad s_{t+i+1|t} =  As_{t+i|t} + Bv_{t+i|t}, \IEEEyessubnumber\\
\quad s_{t+i+1|t} \in \mathcal{X}  \ominus \Omega, \IEEEyessubnumber\\
\quad v_{t+i|t} \in \mathcal{U} \ominus K\Omega, \IEEEyessubnumber\\
\quad s_{t|t} = x_t , \IEEEyessubnumber\\
\quad i = 0,1,...,N-1.\IEEEyessubnumber  
\end{IEEEeqnarray}
    \item State feedback control layer: employs the auxiliary state feedback control law (\ref{eq:robust control law}) 
   that regulates the feedforward term $v_{t|t}^\ast(s_t)$ such that the closed-loop error $e_t$ is bounded inside $\Omega$, which guarantees hard constraint satisfaction of (\ref{eq:robust state constraints}) - (\ref{eq:robust control constraints}).
\end{enumerate}
\begin{remark}
\label{recursive feasibility of RMPC}
The above tube-based RMPC algorithm is  often called closed-loop (CL) MPC, since the initial state $s_{t|t} = x_t$ is the measured state $x_t$ of the system  \cite{mayne2007}\cite{mayne2005}\cite{villa2017}. However, due to disturbances, CL-MPC is not guaranteed to be recursively feasible (i.e. if the OCP problem is feasible at $t=0$, it will remain feasible for all future time steps). 
One way to deal with recursive feasibility is to use $s_{t|t} = x_{t|t}$ whenever the OCP problem (\ref{eq:RMPC}) is feasible, which is known as Mode 1. In case of infeasibility, we switch to a backup control strategy (Mode 2), where we use $s_{t|t} = s_{t+1|t-1}$, namely the current state from the previously optimized feasible trajectory \cite{hewing2018stochastic}. In this case, recursive feasibility is guaranteed, and the resulting RMPC is not a purely state-feedback, but a feedback controller comprising an extended state based on feasibility 
i.e. $u_{t+i|t} = v_{t+i|t}(x_t, s_{t+1|t-1}) + K(x_{t+i|t} - s_{t+i|t}$).  
\end{remark}{}
\section{Stochastic MPC With State And Control Chance Constraints (SMPC)}
\label{sec:SMPC}

The main objectives of SMPC are to deal with computationally tractable stochastic uncertainty propagation for cost function evaluation, and to account for chance constraints, 
where constraints are expected to be satisfied within a desired probability level. With an abuse of notation, we will use some of the notations defined in Section~\ref{sec:RMPC} in a stochastic setting.
\subsection{Stochastic (OCP) formulation and control objectives}
\label{subsec:stochastic problem}
Consider the following discrete-LTI prediction model subject to additive stochastic disturbance $w_t$:
\begin{IEEEeqnarray}{LLL} 
\IEEEyesnumber
\label{eq:stochastic ocp}
x_{t+i+1|t} = Ax_{t+i|t} + Bu_{t+i|t} + w_{t+i}, \IEEEyessubnumber 
\label{eq:stochastic disturbed dynamics}\\
\text{Pr}[H_j x_{t+i+1|t} \leq h_j] \geq 1- \beta_{x_j}, \,\,\, j=1, 2, ..., n_x  \IEEEyessubnumber
\label{eq:state chance constraints}\\
\quad \text{Pr}[G_j u_{t+i|t} \leq g_j] \geq 1- \beta_{u_j}, \,\,\, j=1, 2, ..., n_u  \IEEEyessubnumber
\label{eq:control chance constraints}
\end{IEEEeqnarray}
\begin{assumption}{(Stochastic disturbance)}
\label{ass:stochastic white normal noise}
$w_{t+i} \sim \mathcal{N}(0, \Sigma_w)$ for $i = 0,1,2, ...$ is a disturbance realization of identically independent distributed (i.i.d.), zero mean random variables with normal distribution $\mathcal{N}$. The disturbance covariance $\Sigma_w \in \mathbb{R}^{n\times n} = \text{diag}(\sigma^2_w)$\footnote{$\sigma^2_w \in \mathbb{R}^n = \begin{bmatrix}\sigma^2_1, \sigma^2_2, ..., \sigma^2_n\end{bmatrix}^\top$ denotes the element-wise square operator of the standard deviation vector $\sigma_w$.} is a diagonal matrix, with $\sigma_w \in \mathbb{R}^n$.
\end{assumption}
Eq.~ (\ref{eq:state chance constraints})/(\ref{eq:control chance constraints}) denote individual point-wise (i.e. independent at each point in time) linear state/control chance constraints with a maximum probability of constraint violation $\beta_{x_j}$/$\beta_{u_j}$. Since the disturbed state $x_t$ in (\ref{eq:stochastic disturbed dynamics}) is now a stochastic variable, it is common to split its dynamics $x_{t+i|t} = s_{t+i|t} + \,e_{t+i|t}$ into two terms: a deterministic term $s_{t+i|t} = \mathbb{E}[x_{t+i|t}]$; and a zero-mean stochastic error term $e_{t+i|t} \sim \mathcal{N}(0, \Sigma_{x_{t+i|t}})$, which evolve as
\begin{IEEEeqnarray}{LLL}
\IEEEyesnumber
s_{t+i+1|t} &= A s_{t+i|t} + Bv_{t+i|t}, \quad s_{t|t} = {x}_t
\IEEEyessubnumber
\label{eq:deterministic dynamics}\\
e_{t+i+1|t} &= A_K e_{t+i|t} + w_{t+i}, \quad\,\, e_{t|t} = 0.
\IEEEyessubnumber
\label{eq:stochastic error dynamics}
\end{IEEEeqnarray}
Notice that in contrast to the closed-loop error evolution in RMPC (\ref{eq:robust error dynamics}), the propagation of the predicted error $e_{t+i|t}$ in SMPC is independent of $x_{t+i|t}$ due to the assumption of zero initial error, which enables a closed-loop approach. The evolution of the state covariance 
\begin{IEEEeqnarray}{LLL}
\IEEEyesnumber
\label{eq:state covariance dynamics}
\Sigma_{x_{t+i+1|t}} =  A_K\Sigma_{x_{t+i|t}}A_K^\top + \Sigma_{w}, \quad \Sigma_{x_{t|t}} = 0 
\end{IEEEeqnarray}
is independent of the control. In the same spirit as \cite{Lorenzen2017}\cite{mesbah2017}, the control objective is to bound the stochastic predicted error by employing the following control law:
\begin{IEEEeqnarray}{LLL} 
\label{eq:stochastic control law}
u_{t+i|t} = v_{t+i|t}(x_{t}) + K(x_{t+i|t}-s_{t+i|t}).
\end{IEEEeqnarray}
$K \in \mathbb{R}^{m \times n}$ is a fixed stabilizing dead-beat feedback gains (see remark \ref{remark:choice of feedback gains}) for (\ref{eq:stochastic disturbed dynamics}), and $v_{t+i|t}$ is the decision variable of the MPC program. In what follows, we present a deterministic reformulation of the individual chance constraints (\ref{eq:state chance constraints}) - (\ref{eq:control chance constraints}) that will be used in the SMPC algorithm.

\subsection{Chance constraints back-off design}
\label{subsec:stochastic state back-offs}
 Using the knowledge of the statistics of $x_{t+i|t}$ in (\ref{eq:deterministic dynamics}) - (\ref{eq:stochastic error dynamics}),  individual state chance constraints  can be rewritten as:
\begin{IEEEeqnarray}{LLL} 
\text{Pr}[H_j s_{t+i+1|t}\leq h_j - H_je_{t+i+1|t}] \geq 1- \beta_{j_x}. \IEEEyesnumber
\label{eq:reformulated state chance constraint}
\end{IEEEeqnarray}
We seek the least conservative deterministic  upper bound $\eta_{x_{j,t+i+1|t}}$ such that by imposing
\begin{IEEEeqnarray}{LLL} 
H_j s_{t+i+1|t} \leq h_j -\eta_{x_{j,t+i+1|t}},  \nonumber
\end{IEEEeqnarray} 
we can guarantee that (\ref{eq:reformulated state chance constraint}) be satisfied. The smallest bound $\eta_{x_{j,t+i+1|t}}$ can then be obtained by solving $n_x N$ linear independent chance-constrained optimization problems offline:
\begin{IEEEeqnarray}{LLL} 
\IEEEyesnumber
\label{eq:chance constraint optimization}
\eta_{x_{j,t+i+1|t}} = & \min_{\eta_x}  \quad \eta_x  \\
& \st \quad \text{Pr}[H_je_{t+i+1|t} \le \eta_x] \geq 1- \beta_{x_j}. \nonumber
\end{IEEEeqnarray} 
Using the disturbance assumption (\ref{ass:stochastic white normal noise}), one can solve such programs easily since there exist a numerical approximation of the cumulative density function (CDF) $\phi(\eta_{x_{j,t+i+1|t}}) \geq 1-\beta_{x_j}$ for normal distribution. Hence,  $\eta_{x_{j,t+i+1|t}}$ can be  computed using the inverse of the CDF $\phi^{-1}(1-\beta_{x_j})$ of the random variable $ H_je_{t+i+1|t}$. Contrary to RMPC, the state back-offs grow contractively along the horizon, taking into account the predicted evolution of the error covariance. Similarly, we reformulate the individual control chance constraints in (\ref{eq:control chance constraints}) using  (\ref{eq:deterministic dynamics})-(\ref{eq:stochastic error dynamics}), and the control law (\ref{eq:stochastic control law}):
\begin{IEEEeqnarray}{LLL} 
\text{Pr}[G_j v_{t+i|t}\leq g_j - G_jKe_{t+i|t}] \geq 1- \beta_{u_j}. \IEEEyesnumber
\label{eq:reformulated control chance constraint}
\end{IEEEeqnarray}
 The individual control constraints back-off magnitudes $\eta_{u_{j,t+i|t}}$ can be computed along the horizon using the inverse CDF $\phi^{-1}(1-\beta_{u_j})$ of the random variable $G_jKe_{t+i|t}$. 
\subsection{SMPC with chance constraints algorithm}

The SMPC scheme with individual chance constraints computes feasible reference control actions $\mathbf{v}^\ast(x_t)$ at every MPC sampling time $t$ subject to individual state and control backed-off constraints  as follows
\label{subsec:SMPC algorithm}
\begin{IEEEeqnarray}{LCL}
\label{eq:SMPC}  \IEEEyesnumber
\min_{\mathbf{v}} \quad  \mathbb{E}[J_N(x_t,\mathbf{v})] = (\ref{eq:cost_function}) \IEEEyessubnumber
\label{eq:expected cost fucntion}\\
\st \nonumber \\ \,\, s_{t+i+1|t} =  As_{t+i|t} + Bv_{t+i|t}, \IEEEyessubnumber\\
\,\, H_js_{t+i+1|t} \leq h_j-\eta_{x_{t+i+1|t}}, \, j = 0,1,...,n_x \IEEEyessubnumber\\
\,\, G_jv_{t+i|t} \leq g_j-\eta_{u_{t+i|t}}, \quad\quad j = 0,1,...,n_u \IEEEyessubnumber\\
\,\, s_{t|t} = x_t , 
\label{eq:smpc reintialization}
\IEEEyessubnumber\\
\,\, i = 0,1,...,N-1. \IEEEyessubnumber
\end{IEEEeqnarray}
Note that since the above SMPC algorithm works purely with state-feedback ($s_{t|t} = x_t$), The linear feedback term in ~\eqref{eq:stochastic control law} is only used to predict the variance of the future error $e_t$. 



\begin{remark}
\label{remark:recursive feasibility of SMPC}
The above CL-SMPC algorithm is not guaranteed to be recursively feasible due to the fact that the disturbance realization $w_{t+i} \sim \mathcal{N}(0, \Sigma_w)$ is  unbounded. To tackle this practically, disturbance realizations $w_{t+i}$ are assumed to have a bounded support $\mathcal{W}$ \cite{MAYNE20151}. There have been recent efforts on recursive feasibility for SMPC using different ingredients of cost functions, constraint tightening and terminal constraints as in \cite{Lorenzen2017} \cite{PAULSON2019}. However, recursive feasibility guarantees for SMPC is out of this paper's scope.  
\end{remark}{}

\section{Worst-case Robustness of SMPC}
\label{sec:analysis}

SMPC ensures constraint satisfaction with a certain probability, while RMPC ensures it under bounded disturbances. When comparing the two approaches, one could think that SMPC is equivalent to bounding stochastic disturbances inside a confidence set and then applying RMPC. This section clarifies that this is not the case.
In particular, we answer the following question: when using SMPC, what are the bounded disturbance sets under which we can still guarantee constraint satisfaction? Considering a single inequality constraint and hyper-rectangle disturbance sets, we show how to compute the size of such sets, and that they shrink along the control horizon. Since the disturbance set is instead fixed in RMPC, we conclude that the two approaches are fundamentally different. 

Consider an individual chance constraint \mbox{$\text{Pr}[q^\top x_{t+i+1|t} \leq g] \geq 1 - \beta$}, where $q \in \mathbb{R}^{n}$, $g \in \mathbb{R}$.
Given the corresponding back-off magnitude $\eta_{t+i+1|t}$ \eqref{eq:chance constraint optimization}, we seek the maximum hyper-rectangle disturbance set \mbox{$\mathbb{W}_{t+i} \subset \mathbb{R}^n = \{w : \, |w| \leq w_{t+i}^{max}\}$} such that the constraint $q^\top x_{t+i+1|t} \leq g$ is satisfied for any $w \in \mathbb{W}_{t+i}$:

\begin{IEEEeqnarray}{LLL} 
\eta_{t+i+1|t} = & \max_{e} \, q^\top e \IEEEyesnumber
\\
& \st \nonumber \quad e \in \bigoplus^{i}_{j=0} A_K^j\mathbb{W}_{t+i}. \nonumber
\label{eq:disturbance bound}
\end{IEEEeqnarray}
This problem has a simple solution
\begin{IEEEeqnarray}{LLL} 
\label{eq:disturbance bound}
\eta_{t+i+1|t} = \left(\sum^{i}_{j=0}\left|b_j\right|^\top \right)w^\text{max}_{t+i},\IEEEyesnumber
\end{IEEEeqnarray}
where $b_j^\top \triangleq q^\top A_K^j$ and $|.|$ is the element-wise absolute norm. From the SMPC derivation we know that $\eta_{t+i+1|t}$ is computed via the inverse CDF of $q^\top e_{t+i+1|t}$, which returns a value proportional to its standard deviation $\sigma_{t+i+1|t}$. Therefore we can write
\begin{IEEEeqnarray}{LL} 
\label{eq:eta_smpc}
\IEEEyesnumber
\eta_{t+i+1|t} =  \kappa(\beta) \underbrace{\sqrt{\sum^{i}_{j=0} b_j^\top \Sigma_{w} b_j}}_{\sigma_{t+i+1|t}},
\end{IEEEeqnarray} 
where $\kappa(\beta)$ is a coefficient that depends nonlinearly on $\beta$. By substituting \eqref{eq:disturbance bound} in \eqref{eq:eta_smpc} and exploiting the fact that \mbox{$\Sigma_w = \text{diag}(\sigma_w^2)$} \footnote{$\sigma^2_w \in \mathbb{R}^n = \begin{bmatrix}\sigma^2_1, \sigma^2_2, ..., \sigma^2_n\end{bmatrix}^\top$ denotes the element-wise square operator of the standard deviation vector $\sigma_w$.} $\in \mathbb{R}^{n\times n}$, we infer
\begin{IEEEeqnarray}{LLL} 
\IEEEyesnumber
\label{eq:whatever}
\kappa^2(\beta) \sum^{i}_{j=0}b_j^\top \,\text{diag}(b_j)\,\sigma^2_w = (\sum^{i}_{j=0}| b_j|^\top w^\text{max}_{t+i})^2.
\end{IEEEeqnarray}
Solving for $w^\text{max}_{t+i}$ has infinitely many solutions. However, we can get a unique solution by imposing a ratio $\zeta_{t+i} \in \mathbb{R}$ between $w^\text{max}_{t+i}$ and $\sigma_w$ as follows:
\begin{IEEEeqnarray}{LLL} 
\IEEEyesnumber
w^\text{max}_{t+i} = \zeta_{t+i}\, \sigma_w.
\end{IEEEeqnarray}
Substituting back in (\ref{eq:whatever}) and solving for $\zeta_{t+i}$ we get:
\begin{IEEEeqnarray}{LLL} 
\IEEEyesnumber
\zeta_{t+i} = \kappa(\beta) \sqrt{\alpha_i}, \qquad \alpha_i \triangleq \frac{\sum_{j=0}^i b_j^\top \,\text{diag}(b_j)\,\sigma^2_w}{( \sum_{j=0}^i |b_j|^\top \sigma_w)^{2}}.
\end{IEEEeqnarray}
The series $\alpha_i$ is bounded $0 < \alpha_i \le 1, \quad \forall i\geq 0$, since the sum of squares (numerator) is less than or equal to the square of the sum of positive numbers (denominator). In Appendix \ref{proof}, we prove that $\alpha_i$ is monotonically decreasing (i.e. \mbox{$\alpha_{i+1} < \alpha_i$}) for the case of 1D systems ($n=1$). We confirmed this result numerically for the multi-variate case by randomly generating schur stable closed-loop matrices $A+BK$ subject to the same covariance of the disturbance $\Sigma_w$ for fairness. Since $\alpha_i$ is bounded and monotonically decreasing, then it is convergent. 
This implies that, as $i$ grows, $\zeta_{t+i}$ decreases, and so does the disturbance set $\mathbb{W}_{t+i}$ until it converges in the limit. We conclude that, when using SMPC, the disturbance sets for which we have guaranteed constraint satisfaction shrink along the control horizon.
\section{Simulation Results}
\label{sec:results}
\begin{table}[tbp]
\caption{Modelling and simulation parameters.}
\label{tab:parameters}
\centering
\renewcommand{\arraystretch}{1.5}
\begin{tabular}{|c|c|}
\hline
CoM height ($h$)  & $0.88$ ($m$)\\ \hline
gravity acceleration ($g^z$)  & $9.81$ ($m/s^2$)         \\ \hline
foot support polygon along $x$ direction ($\mathcal{U}^x$) & $\begin{bsmallmatrix}-0.05, & 0.10\end{bsmallmatrix}$ ($m$)        \\ \hline
foot support polygon along $y$ direction ($\mathcal{U}^y$) & $\begin{bsmallmatrix}-0.05, & 0.05\end{bsmallmatrix}$ ($m$)         \\ \hline
bounded disturbance on CoM position ($\mathcal{W}_c$)   & $\begin{bsmallmatrix}-0.0016, & 0.0016\end{bsmallmatrix}$ ($m$)         \\ \hline
bounded disturbance on CoM velocity ($\mathcal{W}_{\dot{c}}$)   & $\begin{bsmallmatrix}-0.016, & 0.016\end{bsmallmatrix}$ ($m/s$)        \\
\hline
disturbance std-dev of CoM position ($\sigma_c$) & $0.0008$ ($m$) \\ \hline
disturbance std-dev of CoM velocity ($\sigma_{\dot{c}}$)  & $0.008$ ($m/s$) \\ \hline
MPC sampling time ($\Delta{t}$) & $0.1$ ($s$) \\ \hline
whole-body tracking controller sampling time & $0.002$ ($s$)  \\ \hline
MPC receding horizon ($N$)  & $16$  \\ \hline
\end{tabular}
\vspace{-0.6cm}
\end{table}

In this section, we present simulation results of the generated walking motions of a Talos robot \cite{Stasse2017} subject to additive persistent disturbances on the lateral CoM dynamics. We compare the motions generated using SMPC subject to state and control chance constraints against nominal MPC and tube-based RMPC. The lateral CoM position is constrained inside a box $-0.04 \leq c^y \leq 0.04$ to avoid collision of the external parts of the robot with walls as it navigates through a narrow hallway with fixed contact locations as shown in Fig. \ref{fig:whole-body motions}. The CoM trajectories generated using MPC are tracked with a Task-Space Inverse Dynamics (TSID) controller using a hard contact model for generating the control commands~\cite{delprete2016}. We use the Pinocchio library
~\cite{carpentier2019} for the computation of rigid-body dynamics. We show an empirical study comparing robustness w.r.t. performance of SMPC against tube-based RMPC and nominal MPC when subject to the same disturbance realizations. The robot model and simulation parameters are defined in Table~\ref{tab:parameters}.   

\subsection{Hard constraints satisfaction in tube-based RMPC}
\label{subsec:RMPC results}
First, we compute offline the state and control back-off magnitudes to tighten the constraint sets for RMPC.

\begin{figure}[tbp]
  \centering
  \includegraphics[width=0.9\columnwidth]{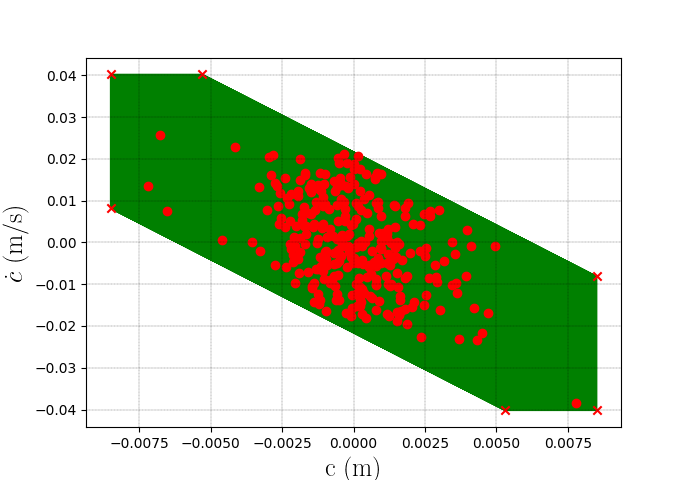}
  \caption{Simulation of $6$ initial conditions (red crosses) at the vertices of the outer-$\epsilon$ approximation of the mRPI set $\Omega$ for $50$ time steps subject to $w_{t+i} \in \mathcal{W}$.}
  \label{fig:mRPI set}
  \vspace{-0.7cm}
\end{figure}

\begin{figure}[tbp]
   \centering
   \includegraphics[width=\columnwidth]{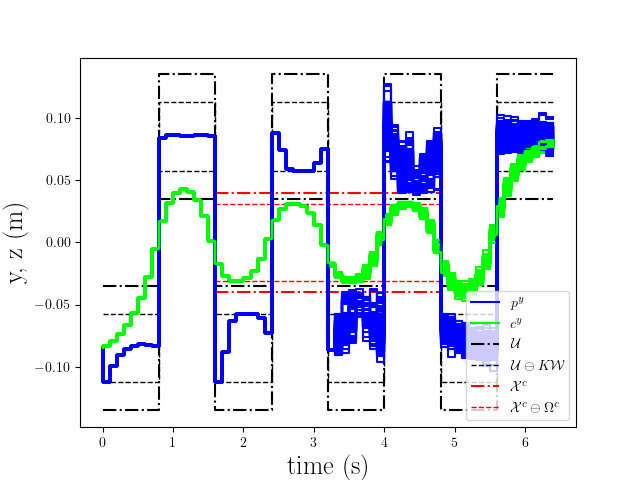}
   \caption{200 simulations of tube-based RMPC with $w_{t+i} \in \mathcal{W}$.}
   \label{fig:RMPC}
     \vspace{-0.6cm}
  \end{figure}

The state constraints back-off magnitude is computed using an outer $\epsilon$ approximation of the mRPI set $\Omega$ using the procedure in \cite{rakovic2005}, with an accuracy of $\epsilon = 10^{-6}$. In Fig. \ref{fig:mRPI set}, we test the positive invariance property (\ref{property: positive invariance}) of $\Omega$, by simulating $6$ initial conditions starting at the set vertices for $50$ time steps, and applying randomly sampled disturbance realizations from the disturbance set $\mathcal{W}$. As shown, the evolution of each initial condition (red dots), is kept inside $\Omega$ (the tube section) when subject to disturbance realizations $w_{t+i} \in \mathcal{W} = \begin{bmatrix}\mathcal{W}_c & \mathcal{W}_{\dot{c}}\end{bmatrix}^\top$.   Using the same choice of pre-stabilizing dead-beat gains $K = \begin{bmatrix}3.386 & 0.968
\end{bmatrix}$ as in \cite{villa2017}, the robust control back-off magnitude $K\Omega$ is computed exactly without resorting to numerical approximation $K\Omega = K\mathcal{W} = \begin{bmatrix}-0.0225,& 0.0225\end{bmatrix}$.
In Fig. \ref{fig:RMPC}, we plot the CoM position and CoP of $200$ trajectories obtained using tube-based RMPC. The robot takes the first two steps in place before entering the hallway. In the third and fourth steps, no disturbances were applied showing that the CoM position $c$ trajectories back off conservatively from the constraint bounds with the magnitude of the mRPI set on the CoM position $\Omega_c$. Finally, we randomly apply sampled Gaussian disturbance realizations $w_{t+i} \sim \mathcal{N}(0, \Sigma_w)$ with finite support $\mathcal{W}$, where $\Sigma_{w} = \left[\begin{smallmatrix}\sigma^2_c & 0 \\ 0 & \sigma^2_{\dot{c}}\end{smallmatrix}\right]$, for the rest of the motion, showing that both state and control constraints are satisfied as expected. 
Note that when the worst-case disturbance is persistently applied on one direction, the state constraint is saturated in that direction as expected. This shows that tube-based RMPC  anticipates for a persistent worst-case disturbance to guarantee a hard constraint satisfaction, which is quite conservative and sub-optimal when the nature of the disturbances is stochastic as in this scenario.

\subsection{Chance-constraints satisfaction in SMPC}  
\begin{figure}[tbp]
  \centering
  \begin{subfigure}{\columnwidth}
  \includegraphics[width=\columnwidth-0.8cm]{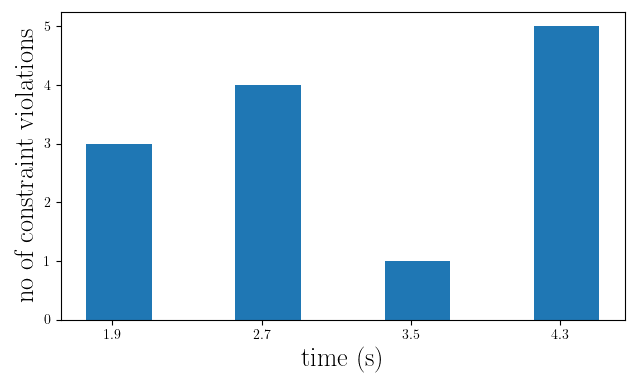}
  \subcaption{CoM position constraint violations.}
  \label{fig:SMPC zoom}
  \end{subfigure}
  \begin{subfigure}{\columnwidth}
  \includegraphics[width=\columnwidth]{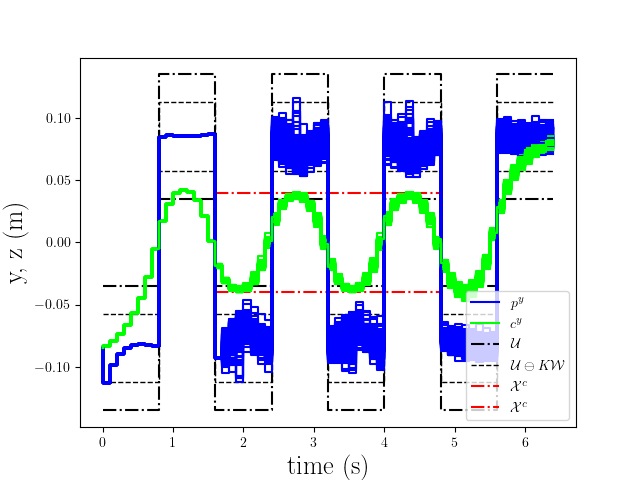}
  \subcaption{CoP and CoM lateral motion with $\beta_{x_j} =\, 5\%$, $\beta_{u_j} =\, 50\%$.}
  \label{fig:SMPC no zoom}
  \end{subfigure}
  \caption{$200$ SMPC simulations with $w_{t+i} \sim \mathcal{N}(0, \Sigma_w) \in \mathcal{W}$.}
  \label{fig:SMPC}
  \vspace{-0.6cm}
  \end{figure}

This subsection presents the results of SMPC. Contrary to  RMPC, the state and control back-off magnitudes ($\eta_{x_{t+i+1|t}}$, $\eta_{u_{t+i}}$) vary along the horizon, and are computed based on the propagation of the predicted state covariance (\ref{eq:state covariance dynamics}), pre-stabilizing feedback gain $K$,  disturbance covariance $\Sigma_{w}$, and the desired probability level of individual state and control constraint violation $\beta_{x_j}$ and $\beta_{u_j}$ respectively. We set $\beta_{x_j} = 5 \%$ , and $\beta_{u_j} = 50\%$, which corresponds to satisfying the nominal CoP constraints. 
Using the same choice of stabilizing feedback gains $K$ as in RMPC, we simulate $200$ trajectories using SMPC in Fig. \ref{fig:SMPC no zoom}. In the first two steps, the robot steps in place and the CoM constraints are not active. For the rest of the motion, we randomly apply sampled Gaussian disturbance
realizations $w_{t+i} \sim \mathcal{N}(0, \Sigma_w)$ with finite support $\mathcal{W}$.
In Fig. \ref{fig:SMPC zoom}, we show the empirical number of CoM position constraint violations out of the $200$ simulated trajectories. The maximum number of constraint violations is obtained at time instance $4.3$s is $5 (\leq 10$), which respects the designed probability level of CoM constraint violations $\beta_{x_j} = 5\%$ as expected. 

\begin{figure}[tbp]
  \centering
  \includegraphics[width=\columnwidth]{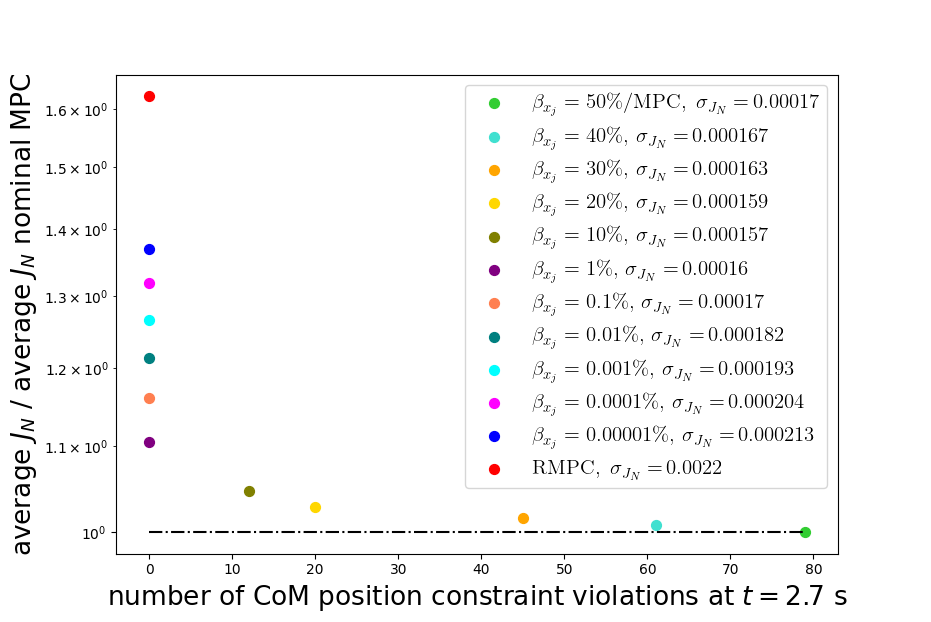}
  \caption{SMPC with varying $\beta_{x_j}$ vs RMPC of $200$ simulations with $w_{t+i} \sim \mathcal{N}(0, \Sigma_w) \in \mathcal{W}$. The dotted line denotes the optimal ratio of one (nominal MPC)}
  \label{fig:violations vs cost}
  \vspace{-0.5 cm}
\end{figure}

To test robustness of constraint satisfaction and optimality of SMPC, we ran an empirical study of the same eight step walking motion ($200$ trajectories) comparing SMPC with varying $\beta_{x_j} \in  \begin{bmatrix} $0.00001\%$,& 50 \%\end{bmatrix}$ and fixed $\beta_{u_j} = 50 \%$ against tube-based RMPC and nominal MPC in Fig. \ref{fig:violations vs cost}. We plot the empirical number of CoM position constraint violations at $t = 2.7$s, against the averaged cost performance (of $200$ trajectories) ratio between different MPC schemes and nominal MPC. As before, disturbance realizations are sampled from $\mathcal{N}(0, \Sigma_w)$ with finite support $ \mathcal{W}$. As expected, the higher the probability level of constraint satisfaction in SMPC, the lower the amount of constraint violations (higher robustness). The highest number of constraint violations is obtained at $\beta_{x_j} = 50 \%$, which is equivalent to nominal MPC. Zero constraint violations were obtained when $\beta_{x_j} \leq 1 \%$, as for RMPC. An advantage of SMPC  with $\beta_{x_j} \leq 1 \%$ over RMPC, is the lower average cost. This gives the user the flexibility to design the controller for different task constraints, by tuning the  probability level of constraint satisfaction without sacrificing performance as in tube-based RMPC or sacrificing robustness as in nominal MPC. 

\section{Discussion and Conclusions}

This paper compared the use of SMPC with RMPC to account for uncertainties in bipedal locomotion. Many SMPC and RMPC algorithms exist. We decided to focus on two particular instances of tube-based approaches, which have the same online computational complexity as nominal MPC. Indeed, all the extra computation takes place offline, and consists in the design of tightened constraints (back-offs) based on a fixed pre-stabilizing feedback gain $K$. 

Our comparison focused on the trade off between robustness and optimality. 
Our tests show that, while SMPC does not provide hard guarantees on constraint satisfaction, in practice we did not observe any constraint violation with sufficiently low $\beta_x (\leq 1\%)$. This comes with the advantage of less conservative control, i.e. it results in better performance as measured by the cost function. This is reasonable because RMPC behaves conservatively, expecting a persistent worst-case disturbance, which in practice is extremely unlikely to happen. SMPC instead reasons about the probability of disturbances. In Section~(\ref{sec:analysis}) we showed that we can compute the maximum disturbance sets to which SMPC ensures robustness. We showed that these sets shrink contractively as time grows. Loosely speaking, SMPC can be thought as a special kind of RMPC that considers shrinking disturbance sets along the horizon.

Our empirical results are specific to the choice of dead-beat feedback gains used in both algorithms. These gains were computed in \cite{villa2017} by minimizing the back-off magnitude on the CoP constraints. This is sensible because the CoP is usually more constrained than the CoM in bipedal locomotion. Other feedback gains could be used, such as LQR gains, resulting in back-off magnitudes that are a trade-off between state and control constraints. 
While changing the gains would affect our quantitative results, it would not affect the qualitative differences between SMPC and RMPC that we highlighted in the paper.

In conclusion, SMPC offers an opportunity for the control of walking robots that affords trading-off robustness to uncertainty and performance.
For Future work, we intend to investigate nonlinear versions of RMPC and SMPC~\cite{koehler2019},\cite{santos2019} to enable the use of more complex models of locomotion.

\appendix
\begin{figure*}
\centering
\includegraphics[width=0.19\textwidth]{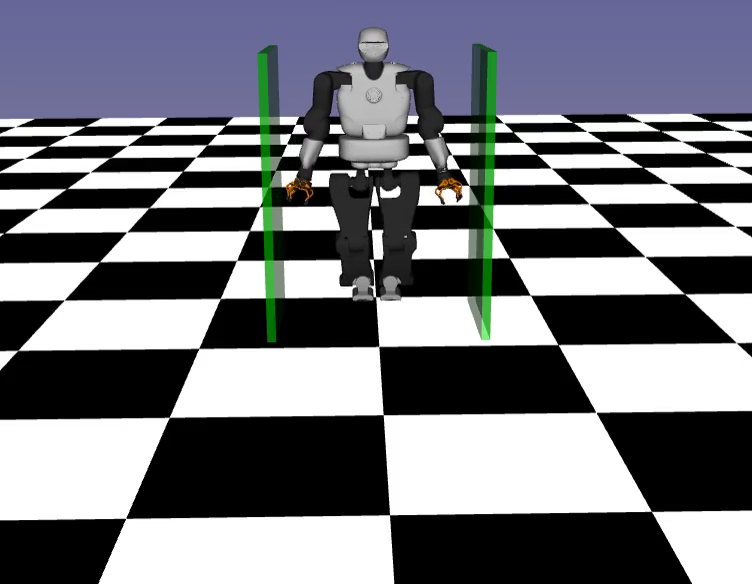} 
\includegraphics[width=0.19\textwidth]{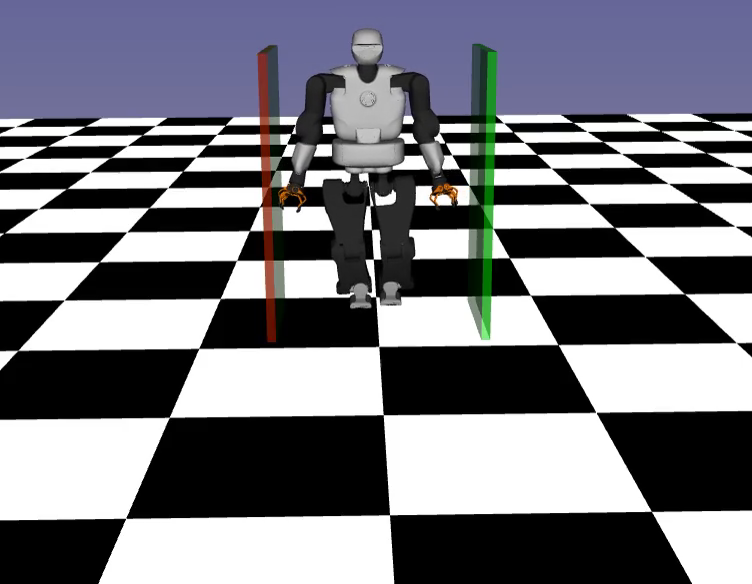}
\includegraphics[width=0.19\textwidth]{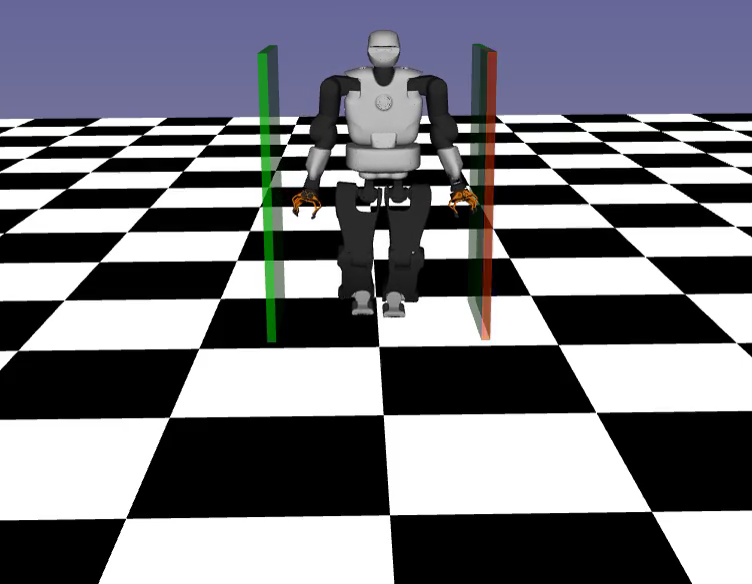}
\includegraphics[width=0.19\textwidth]{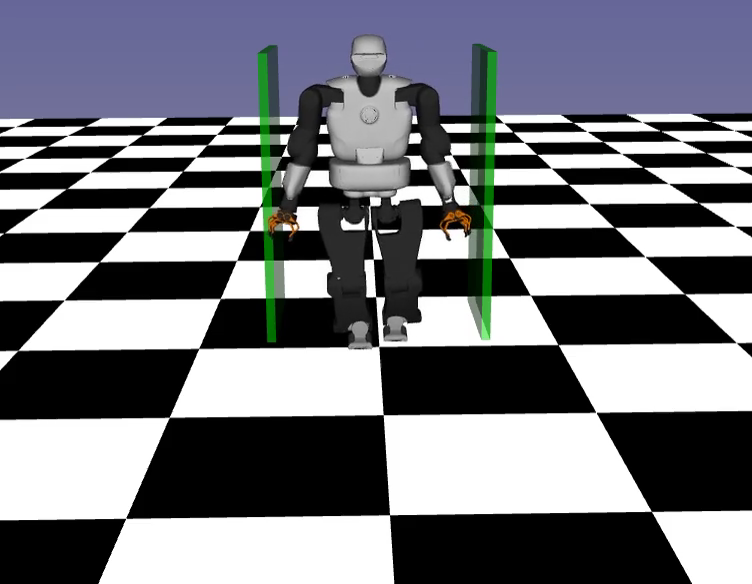}
\includegraphics[width=0.19\textwidth]{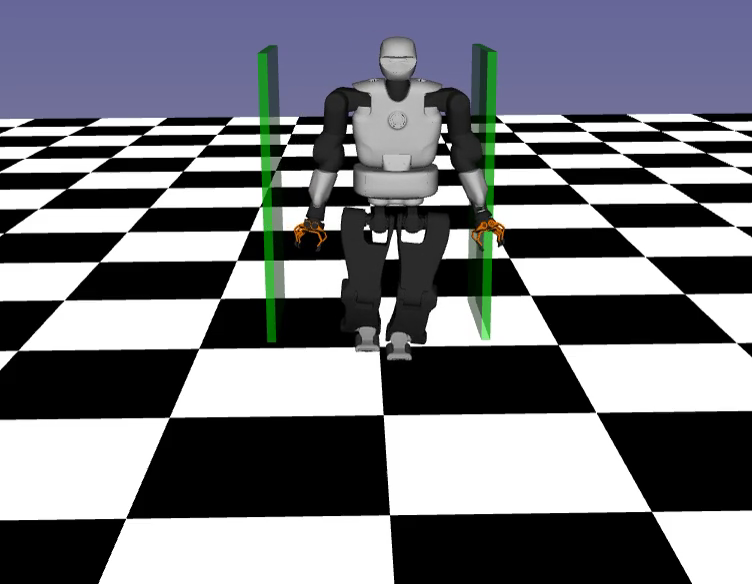}\\
\caption{TALOS robot walking through a narrow hallway using nominal MPC subject to additive disturbances on the lateral CoM dynamics. The Red color corresponds to the robot colliding with the wall.} 
\label{fig:whole-body motions}
  \vspace{-0.6cm}   
\end{figure*}
\renewcommand{\qedsymbol}{\rule{0.7em}{0.7em}}
\subsection{Proof that $\alpha_i$ is monotonically decreasing (1D case):}
\label{proof}
We would like to show that $\alpha_{i+1} < \alpha_i\, , \,\, \forall i\geq0$. Given 
\begin{IEEEeqnarray}{LLL}
\alpha_{i} = \frac{\sum_{j=0}^i b^2_j\,\sigma^2_w}{( \sum_{j=0}^i |b_j|\sigma_w)^{2}} \quad \alpha_{i+1} = \frac{\sum_{j=0}^{i+1} b^2_j\,\sigma^2_w}{(\sum_{j=0}^{i+1} |b_j|\sigma_w)^{2}}, 
\end{IEEEeqnarray}
where $b_j = q^\top A_K^j$ can be written as $q a^j$, with \mbox{$a\triangleq A_K$}, \mbox{$|a| <1$}. After simplifying $\sigma_w$ we get:
\begin{IEEEeqnarray}{LLL}
\label{ineq: main inequality}
\scalebox{1.0}[1.0]{$
 \frac{\sum_{j=0}^{i} b^2_j + b^2_{i+1}}{(\sum_{j=0}^{i} |b_j|)^2 + b^2_{i+1} + 2|b_{i+1}|\sum_{j=0}^{i} |b_j|} \le \frac{\sum_{j=0}^i b^2_j}{( \sum_{j=0}^i |b_j|)^2}.$}
\end{IEEEeqnarray}
By substituting the analytical expressions of the following series in (\ref{ineq: main inequality})
\begin{IEEEeqnarray}{LLL}
\small
\IEEEyessubnumber
\small
\sum_{j=0}^{i} |b_j| = |q|\sum_{j=0}^{i}|a|^j = |q| 
\left(\frac{1-|a|^{i+1}}{1-|a|}\right),\\ 
\IEEEyessubnumber
\sum_{j=0}^{i} b^2_j = q^2\sum_{j=0}^{i}a^{2j} = q^2 
\left(\frac{1-a^{2(i+1)}}{1-a^2}\right).
\end{IEEEeqnarray}
and cross multiplication, we get
\begin{IEEEeqnarray}{LLL}
\label{ineq:expanded inequality}
\small |q|^3\left(\frac{1-|a|^{i+1}}{1-|a|}\right)^2 |a|^{i+1} \leq \nonumber\\  q^2\left(\frac{1-a^{2(i+1)}}{1-a^2}\right)\left(|q a^{i+1}| + 2|q| \left(\frac{1-|a|^{i+1}}{1-|a|}\right)\right).
\end{IEEEeqnarray}
By multiplying both sides of \eqref{ineq:expanded inequality} by $\frac{(1-|a|)^2}{|q|^3}$, we have
\small 
\begin{IEEEeqnarray}{LLL}
\small |a|^{i+1}(1-|a|^{i+1})^2 \leq \frac{1-a^{2(i+1)}}{1+a} \left(|a|^{i+1}(1-|a|) +2 (1-|a|^{i+1})\right) \nonumber \\
\Rightarrow (1+|a|)(|a|^{i+1}-a^{2i+2}) \leq (1+|a|^{i+1}) (-|a|^{i+1}-|a|^{i+2} + 2 ) \nonumber \\
\Rightarrow |a|^{i+1} + |a|^{i+2} - |a|^{2i+3} \leq 2 + |a|^{i+1}-|a|^{i+2} -|a|^{2i+3} \nonumber \\
\Rightarrow 2 - 2 |a|^{i+2} \ge 0,
\end{IEEEeqnarray}
\normalsize
which always holds because $|a|<1$. This concludes the proof. \qed

\addtolength{\textheight}{0cm}     

\bibliographystyle{IEEEtran}
\bibliography{IEEEabrv,Biblio}

\end{document}